\pdfoutput=1
\documentclass[12pt,titlepage]{article}
\usepackage{verbatim}
\usepackage{url}
\usepackage{hyperref}
\usepackage{cite}
\usepackage{float}
\usepackage{graphicx}
\usepackage[title]{appendix}
\usepackage{epstopdf}
\usepackage{amsmath,amsfonts,amsthm,bm}
\usepackage{layout}
\usepackage{setspace}
\usepackage{caption}
\usepackage[T1]{fontenc}
\usepackage[utf8]{inputenc}
\usepackage{authblk}
\DeclareUnicodeCharacter{202F}{\,}
\title{\bf Estimation of Residential Radon Concentration in Pennsylvania Counties by Data Fusion}
\author[1]{Xuze Zhang}
\author[2]{Saumyadipta Pyne}
\author[1]{Benjamin Kedem}
\affil[1]{Department of Mathematics and Institute for Systems Research, University of Maryland, College Park}
\affil[2]{Public Health Dynamics Laboratory, and Department of Biostatistics, Graduate School of Public Health,
University of Pittsburgh, Pittsburgh}
\begin{document}

\maketitle

\begin{abstract}
A data fusion method for the estimation of residential radon level distribution in any 
Pennsylvania county is proposed. The method is based on
a multi-sample density ratio model with variable tilts and is applied to combined
radon data from a reference county of interest and its neighboring counties.
Beaver county and its four immediate neighbors are taken as a case in point.
The distribution of radon concentration is estimated in each of six periods,
and then  the analysis is repeated combining the data from all the periods
to obtain estimates of 
Beaver threshold probabilities and the corresponding confidence intervals. 
\end{abstract}

\section{Introduction}

Radon-222, or just radon, is a tasteless, colorless and odorless radioactive gas, which is a product of Uranium-238 and Radium-226, both of which are naturally abundant in the soil. 
Radon is known worldwide as a carcinogen as its inhaled decay products can get trapped in the lung and induce mutations in DNA. It is the leading cause of lung cancer among non-smokers \cite{WHO2009}. 
According to an estimation of the impact of radon-associated lung cancer by the U.S. Environmental Protection Agency (EPA), “out of a total of 157,400 lung cancer deaths nationally in 1995, 21,000 (13.4\%) 
were radon related. Among non-smokers, an estimated 26\% were radon related” \cite{EPA2003}.
In 1988, radon was formally categorized as a Group 1 Carcinogen (Human Carcinogen) by the International Agency for Research on Cancer \cite{WHO2009}. 
Many countries now pay more attention to this challenge not only as an occupational hazard to miners, but also as a serious risk factor to the health of the general population 
resulting from indoor residential exposure via cracks in basement floors, walls, etc. \cite{DHHS2012}.
Building regulations in radon affected areas have been introduced in countries such as the 
United Kingdom. In 1986, the EPA set an action level (the level above which mitigation is recommended) for residential radon to 4  picocuries per liter 
(pCi/L). Yet, approximately 1 in 15 residences in the U.S. may have radon levels higher than the action level. 
Although mitigation of radon can be achieved effectively through ventilation and suitable building materials, elevated radon levels are of concern to buyers and renters and can therefore impact real estate values.

Historically, residences in certain regions of the state of Pennsylvania (PA) have recorded relatively high radon concentrations. For example, the counties that overlie Reading Prong in southeastern PA were found to have elevated residential radon concentrations \cite{CaseyEtAl2015}. Approximately 40 percent of Pennsylvania homes have radon levels above the EPA action guideline of 4 pCi/L \cite{PA.GOV}. In the present study, we used county-level indoor radon concentrations based on records collected between 1989 and 2017 from Beaver county and its neighboring counties in PA. Most of the records were reported to the PA Department of Environmental Protection (PA DEP), Bureau of Radiation Protection, Radon Division, during real estate transactions, as submitted by certified test companies, laboratories, or real estate owners.

In this paper, we show how to estimate the radon concentration distribution in any PA
county of interest by combining or fusing
the county's radon data with the data from its neighboring counties, using the so called  
{\em density ratio model} (DRM) with {\em variable tilts}. 
Such flexibility mitigates  the problem of misspecified DRM, and is a new feature of the paper.

Our data consist of radon concentration in Pennsylvania's 67  counties over 6 periods,
but as a case in point 
we shall focus on Beaver County and its neighboring counties,
Lawrence, Butler, Allegheny, and Washington. We shall refer to the county of interest,
Beaver in the present case, as the {\em reference} county.

We follow a two-stage procedure.
For each reference county 
we first estimate its radon concentration distribution for each period by DRM,
using the combined data from the reference county and its neighbors.
In the present case of Beaver and its neighbors, 
as the radon distributions in each of the  six periods behave similarly,
we proceeded next to combine the data from the different periods and used the DRM again to get an improved estimate of the  radon concentration distribution for the 
reference county (Beaver).
Unlike the present case, when the radon distributions in each of the  six periods behave differently
we use the results from the last period only. As mentioned, the entire analysis is based on DRM with variable tilts, a novel idea to be clarified in what follows.

\section{Methodology}

\subsection{Data Description}

\begin{figure}[htbp]
\begin{center}
\includegraphics[width=6in]{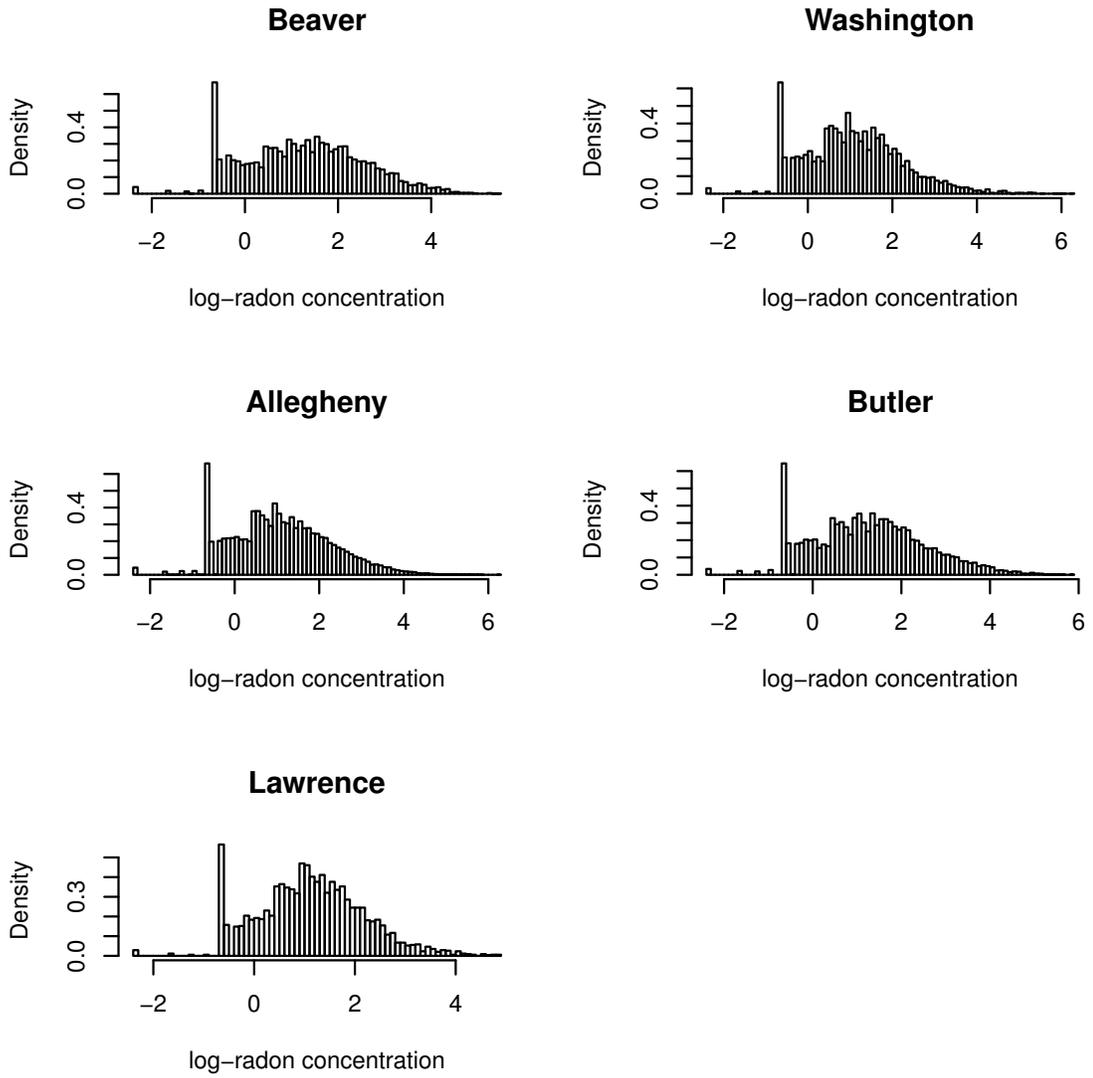}
\caption{Typical histograms of log-radon concentration from five Pennsylvania counties.}
\label{Hist 6 counties}
\end{center}
\end{figure}

\begin{figure}[htbp]
\begin{center}
\includegraphics[width=6in]{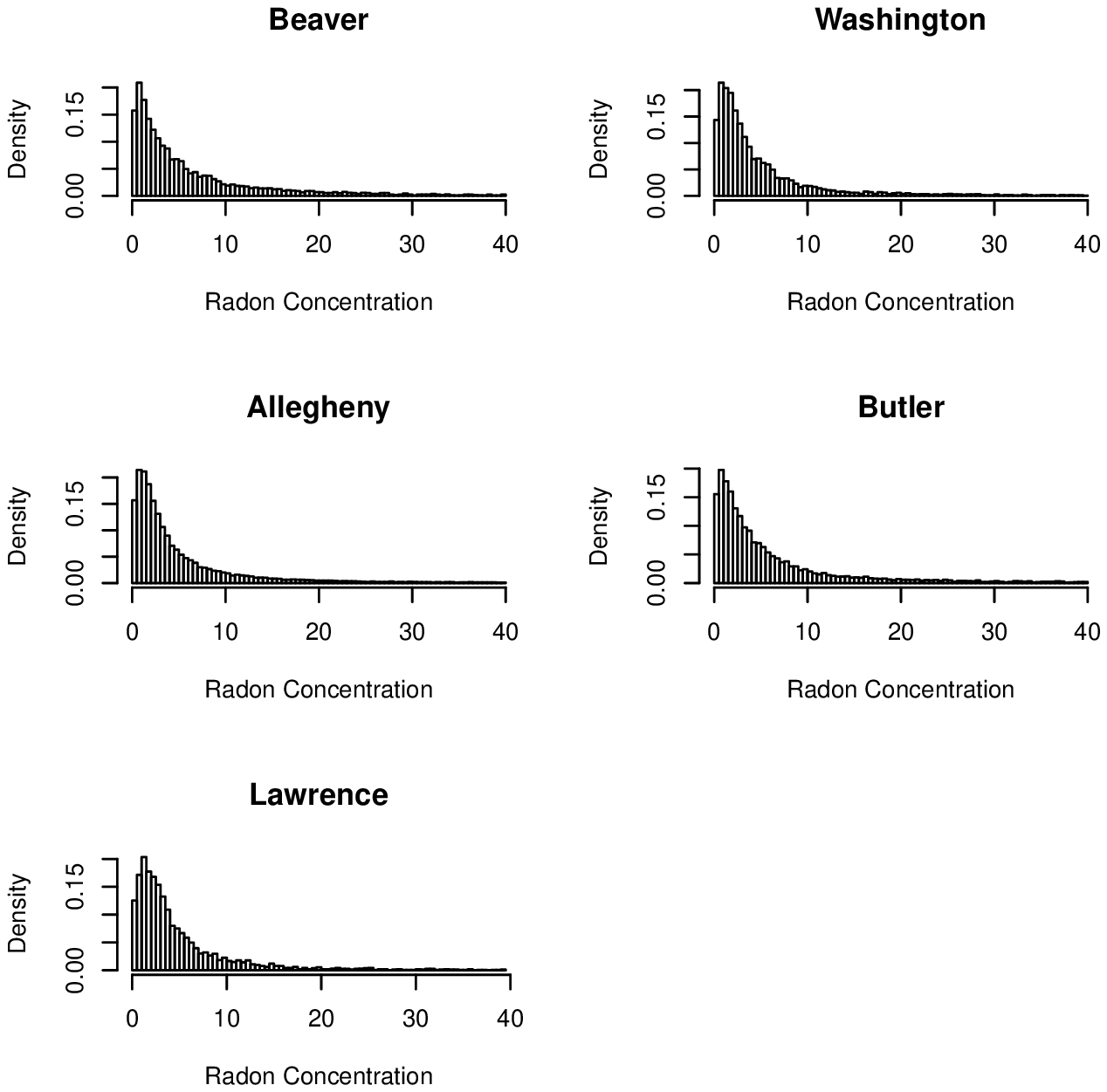}
\caption{Typical histograms of radon concentration from five Pennsylvania counties
truncated at 40 pCi/L.}
\label{Truncated Hist 6 counties}
\end{center}
\end{figure}

The following histograms 
from five Pennsylvania counties depict 
typical distributions of radon concentration. 
For a better visualization, the histograms of log-radon concentration are shown in Figure \ref{Hist 6 counties}. To discern the distribution pattern near 0, the histograms in  
Figure \ref{Truncated Hist 6 counties} 
are constructed from  
observations less than 40 pCi/L.

From the histograms in Figures \ref{Hist 6 counties} and \ref{Truncated Hist 6 counties} 
it is observed that the data are positive and right skewed. 
Experience shows that 
both the 
\emph{Gamma} and \emph{Lognormal} distributions are possible models for 
positive right skewed data. 
However, as the tail behavior is sensitive to distribution assumptions, either choice
could lead to a 
misspecified model. 
To alleviate this problem, we render the distribution assumption less strict via a 
density ratio hybrid structure 
which caters to both distribution models simultaneously. 
A goodness-of-fit plot in Section \ref{GoodnessOfFit}
supports this procedure in the present case of radon data.
This graphical goodness-of-fit has been introduced in \cite{Voulgaraki2011},

\subsection{Density Ratio Model}

Instead of making particular
assumptions about the distributions of radon concentration in each county, 
we shall estimate the distributions using 
the following DRM  
\cite{FKQS},\cite{Fokianos2004},\cite{SDF},\cite{QinZhang97}\cite{QinLawless},
\begin{equation}
\frac{g_{k}(x)}{g_{0}(x)}=\exp{(\alpha_k+\bm{\beta}^{T}_{k}\bm{h}_{k}(x))} \phantom{aaaaa}     k=1, \dots, m
\label{DRM}
\end{equation}
where $g_{0}(x)$ is the reference radon pdf, $g_{k}(x)$'s are the pdf's of radon 
concentration in the neighboring counties, and $m$ denotes the number of neighboring counties. 
The tilt functions
$\bm{h}_{k}(x)$'s with respective dimensions $r_1, \dots, r_m$ describe the density ratio structure.

We shall first assume the {\em uniform} or {\em global} tilt
$\bm{h}_{k}(x)=(x, \log(x), \log^{2}(x))^T$ for all $k$. Semiparametric inference about the densities and the parameters is then obtained by
 combining the data of the reference county with that of its neighboring counties. 
Any of the tilt components $x, \log(x), \log^{2}(x)$ is removed 
when the corresponding $\beta$ coefficient is not significantly different from zero. In this sense
the density ratio structure is a variable structure, so that different counties could be
represented with different tilts relative to the reference.

The tilt $(x, \log(x), \log^{2}(x))^T$ is parsimonious in the sense that its dimension 
is only 3. However, it has been found  sufficient for describing the density ratio 
structure in the present radon case.

Note that the tilt of two $Gamma$ densities is $(x, \log(x))^T$ and the tilt of two $Lognormal$ 
densities is $(\log(x), \log^{2}(x))^T$ 
so that $(x, \log(x), \log^{2}(x))^T$
can represent the density ratio between two $Gamma$ and $Lognormal$ densities.
Hence, the assumption of  the tilt $(x, \log(x), \log^{2}(x))^T$ is less strict in the sense that it allows the distribution to vary between a thin-tailed distribution and a heavy-tailed distribution.

\subsection{Estimation and Asymptotic Result}

More details regarding 
the theoretical underpinnings of this and the following section are given in
the Appendix.

Suppose $\bm{X}_0=(X_{01}, \dots, X_{0n_0})^T$ is a sample from the county of interest with size $n_0$, and $\bm{X}_1, \dots, \bm{X}_m$ are the samples from the neighboring counties with sizes $n_1, \dots, n_m$, respectively. As above, $\bm{X}_0$ denotes the reference sample, 
and let $G$ be the corresponding reference CDF. Denote the combined or fused sample of
size $n=\sum_{k=0}^{m}n_k$ by 
$\bm{t}=(\bm{X}^{T}_{0}, \dots, \bm{X}^{T}_{m})^T$.

The parameters are estimated by maximizing the following empirical likelihood
\begin{equation}
L(\bm{\alpha}, \bm{\beta}, G)=\prod_{i=1}^{n}p_i\prod_{k=1}^{m}\prod_{j=1}^{n_k}w_{k}(X_{kj})
\end{equation}
subject to the constraints 
\begin{equation}
\sum_{i=1}^{n}p_i=1 \phantom{aaa}\sum_{i=1}^{n}p_{i}[w_{k}(t_i)-1]=0 \phantom{aaa} k=1, \dots, m
\end{equation}
where $p_i=dG(t_i)$ and $w_k(\cdot)=\exp{(\alpha_k+\bm{\beta}^{T}_{k}\bm{h}_{k}(\cdot))}$. 
Denote the parametric
estimators by $\tilde{\bm{\alpha}}$, $\tilde{\bm{\beta}}$ and $\tilde{p}_i$'s.

As $n\to \infty$, the estimators $\tilde{\bm{\alpha}}$, $\tilde{\bm{\beta}}$ 
are asymptotically normal, 
\begin{equation}
\sqrt{n}\left(
  \begin{array}{c}   
    \tilde{\bm{\alpha}}-\bm{\alpha}_{0}\\  
    \tilde{\bm{\beta}}-\bm{\beta}_{0}\\  
  \end{array}
\right)\stackrel{d}{\to}N(\bm{0}, \bm{\Sigma}) 
\label{AsymptoticNormal}             
\end{equation}
where $\bm{\alpha}_{0}$ and $\bm{\beta}_{0}$ are the true parameters. 
Let $\tilde{G}(t)=\sum_{i=1}^{n}\tilde{p}_{i}I[t_i\leq t]$. Then, following \cite{Lu2007}
we shall show in the Appendix there are $\bm\Sigma$ and $\sigma(t)$ such that,
\begin{equation}
\sqrt{n}(\tilde{G}(t)-G(t))\stackrel{d}{\to}N(0, \sigma(t)).             
\end{equation}

\subsection{Test of Significance and Tilt Choice}

We start with the uniform or global tilts $(x, \log(x), \log^{2}(x))^T$, and then test the significance of the 
component-functions $x, \log(x), \log^{2}(x)$,
since the density ratio structure could vary between different pairs of counties. 
Testing the tilt components requires first  pairwise DRM fitting 
using the combined data from the reference and each of its neighbors.

Suppose we combine the data of the reference county and its $i$'th neighbor,
 such that $\bm{t}=(\bm{X}^{T}_{0}, \bm{X}^{T}_{i})^T$. Based on 
(\ref{AsymptoticNormal}), we have
\begin{equation}
\sqrt{n}(\tilde{\bm{\beta}}_{i}-\bm{\beta}_{i})\stackrel{d}{\to}N(\bm{0}, \bm{\Sigma}_{\bm{\beta}_{i}}),
\end{equation}
and hence we can test the hypothesis $H_0: \bm{\beta}_i=\bm{0}$ by the chi-square test
\begin{equation}
\chi^2=n\tilde{\bm{\beta}}^{T}_{i}\tilde{\bm{\Sigma}}^{-1}_{\bm{\beta}_{i}}\tilde{\bm{\beta}}_{i}\stackrel{d}{\to}\chi^{2}_{\nu}
\end{equation}
where $\nu$ is the dimension of $\bm{\beta}_{i}$. If we do not reject the hypothesis, then 
the reference and its neighbor are equidistributed.
On the other hand,
if the hypothesis is rejected, then we test the significance of each 
$\beta_{ij}$ from $\bm{\beta}_{i}=(\beta_{i1}, \dots, \beta_{ir_i})$ 
using the Z-test defined by
\begin{equation}
Z=\frac{\tilde{\beta}_{ij}}{\sqrt{\frac{\tilde{\sigma}^{2}_{ij}}{n}}}
\end{equation}
where $\tilde{\sigma}^{2}_{ij}$ is the corresponding estimated variance of $\tilde{\beta}_{ij}$. 
If $\beta_{ij}$ is insignificant, we eliminate the corresponding term in the tilt function 
to form a reduced tilt function.

\subsection{Refined Model and Goodness-of-Fit}
\label{GoodnessOfFit}

Once the pairwise tilt functions are refined by the previous preliminary analysis, 
the density ratio model (\ref{DRM})
 is fitted again with the modified tilts, using the combined data from the reference
county and {\em all} its neighbors, either for each period or for all the periods. The
Beaver example in the Section \ref{BeaverExample} clarifies this point.
In that way we 
obtain the estimated reference  CDF $\tilde{G}$ and, for any threshold $T$,
the estimated threshold probabilities $1-\tilde{G}(T)$ and their asymptotic 
confidence intervals.

The validity of the density ratio structure with variable tilts can be verified by plotting 
$\tilde{G}$, an outcome of DRM, versus the model free
empirical distribution function $\hat{G}$ obtained
from the reference sample only. 
Such goodness-of-fit plots are useful  graphical measures which show
the closeness of $\tilde{G}$ to $\hat{G}$ \cite{Voulgaraki2011},\cite{VoulgarakiEtAl2012}.

\section{Simulation}

In this section it is illustrated, by a simulation of a special case, 
that the estimate of the reference CDF obtained by a DRM with variable tilts is 
more precise than both of the estimates obtained by a DRM with a global tilt and by the empirical CDF 
from the reference $\bm{X}_{0}$ in terms of mean integrated absolute error (MIAE) and mean integrated squared error (MISE),
\begin{equation}
\begin{split}
\mathrm{MIAE}&=\mathrm{E}\int{\bigg|\hat{f}(x)-f(x)\bigg| dx}\\
\mathrm{MISE}&=\mathrm{E}\int{\bigg(\hat{f}(x)-f(x)\bigg)^{2}dx}.
\end{split}
\end{equation}
Here $\hat{f}$ represents the three different estimates $\tilde{G}_{u}$, $\tilde{G}_{r}$ and $\hat{G}$, and $f$ is
 the true reference CDF $G$.
 
So,
consider three random samples $\bm{X}_{0}\sim Exp(2)$, $\bm{X}_{1}\sim Gamma(2, 2)$ and $\bm{X}_{3}\sim Lognormal(1, 1)$ with size $n_0$, $n_1$ and $n_2$, respectively. They follow the 
true density ratio structure  
$$\frac{g_{1}(x)}{g_{0}(x)}=\exp(\log2+\log(x))$$
 and 
$$\frac{g_{2}(x)}{g_{0}(x)}=\exp(-\frac{1}{2}-\log2\sqrt{2\pi}+2x-\frac{1}{2}(\log^{2}(x))).$$

We constructed two DRM's, one where both $\frac{g_{1}(x)}{g_{0}(x)}$ and
$\frac{g_{2}(x)}{g_{0}(x)}$ are
with the global (``u") tilt $(x, \log(x), \log^2(x))^T$, 
and the other where 
$\frac{g_{1}(x)}{g_{0}(x)}$ and
$\frac{g_{2}(x)}{g_{0}(x)}$ 
are 
with the refined (``r") tilts 
$\bm{h}_{1}(x)=\log(x)$ and $\bm{h}_{2}(x)=(x, \log^2(x))^T$, respectively.
We then obtained the corresponding estimated reference CDF's $\tilde{G}_{u}$ and $\tilde{G}_{r}$, from the combined data,
as well as the empirical estimate $\hat{G}$ from the reference sample only,
and compared them with true reference CDF $G$.

Thus, we follow the steps:\\
\\
1.\phantom{a}Generate random samples $\bm{X}^{(i)}_{0}\sim Exp(2)$, 
$\bm{X}^{(i)}_{1}\sim Gamma(2, 2)$ and 
\\\phantom{aa} $\bm{X}^{(i)}_{3}\sim Lognormal(1, 1)$ with size $n_0$, $n_1$ and $n_2$.\\
\\
2. Obtain the estimates $\tilde{G}^{(i)}_{u}$, $\tilde{G}^{(i)}_{r}$ and $\hat{G}^{(i)}$.\\
\\
3. Repeat 1 and 2 for $i=1, \dots, I$\\
\\
4. Approximate MIAE and MISE by 
\begin{equation}
\begin{split}
\widehat{\mathrm{MIAE}}&=\frac{1}{I}\sum_{i=1}^{I}\delta\sum_{j=1}^{J}\bigg|\hat{f}(M_1+j\delta)-f(M_1+j\delta)\bigg|\\
\widehat{\mathrm{MISE}}&=\frac{1}{I}\sum_{i=1}^{I}\delta\sum_{j=1}^{J}\bigg(\hat{f}(M_1+j\delta)-f(M_1+j\delta)\bigg)^2
\end{split}
\end{equation}
where $(M_1, M_2)$ is the region we integrate over that satisfies $\int_{-\infty}^{\infty}f(x)dx\approx\int_{M_1}^{M_2}f(x)dx$ and $\delta=\frac{M_2-M_1}{J}$. 

For our simulation
we set $I=300$, $n_0=n_1=n_2=1000$, $(M_1, M_2)=(0, 10)$ and $J=1000$. The results
 shown in Table \ref{SimRes} indicate that the DRM with the refined tilts gives a better estimate of the reference CDF  judging by both measures MIAE and MISE.

\begin{table}[H]
\centering
\captionsetup{margin=10pt,font=small,labelfont=bf}
\caption{MIAE and MISE for $\tilde{G}_{u}$, $\tilde{G}_{r}$ and $\hat{G}$}
\scalebox{0.9}{
\begin{tabular}{cccc}
\hline
Estimate & $\tilde{G}_{u}$ & $\tilde{G}_{r}$ & $\hat{G}$ \\
\hline
MIAE & $1.639\times 10^-2$ & $1.420\times10^-2 $ & $1.910\times10^-2$ \\
MISE & $2.015\times10^-4$ & $1.712\times10^-4$ & $2.410\times10^-4$ \\
\hline 
\end{tabular}
}
\label{SimRes}
\end{table}

\section{Beaver Example}
\label{BeaverExample}

Beaver county has 4 neighboring counties: Washington, Allegheny, Butler and Lawrence. The 
radon sample sizes are shown in Table \ref{Size} .

\begin{table}[H]
\centering
\captionsetup{margin=10pt,font=small,labelfont=bf}
\caption{Sample sizes of Beaver, Washington, Allegheny, Butler and Lawrence}
\scalebox{0.9}{
\begin{tabular}{cccccc}
\hline
& Beaver & Washington & Allegheny & Butler & Lawrence \\
\hline
89-93 & 816 & 471 & 12328 & 985 & 231\\
94-98 & 913 & 632 & 11982 & 2046 & 476 \\
99-03 & 797 & 642 & 6548 & 1192 & 400 \\
04-08 & 1209 & 1017 & 7581 & 1891 & 753\\
09-13 & 2064 & 1820 & 12772 & 3142 & 910\\
14-17 & 1626 & 1472 & 8419 & 1995 & 660\\
Total  & 7425 & 6054 & 59630 & 11251 & 3430\\
\hline 
\end{tabular}
}
\label{Size}
\end{table}

We constructed the density ratio model as in (\ref{DRM}) with the uniform tilt function
$\bm{h}_{k}(x)=(x, \log(x), \log^{2}(x))^T$ for 
$k$=1-Washington, 2-Allegheny, 3-Butler, 4-Lawrence.

For each period, 
we first combined the data from Beaver and each  of its neighbors (i.e. pairwise fusion)
to get in each case a possibly reduced tilt. The possibly reduced tilts are then used in
forming the density ratio model for each period, using the combined data from all four
counties plus the reference Beaver county. In that way each period gives us the 
estimated $\tilde{G}$.
Observe that, for each period, the empirical CDF $\hat{G}$ is obtained from the reference county (Beaver) only and is model free.
The refined tilt functions for each period are given in Table \ref{Refined Tilts}.

\begin{table}[H]
\centering
\captionsetup{margin=10pt,font=small,labelfont=bf}
\caption{Refined tilts for each county relative to Beaver, in each period.
A hyphen ``-" indicates that the radon concentration of the indicated county and that of
Beaver are identical.}
\scalebox{0.9}{
\begin{tabular}{ccccc}
\hline
 & Washington & Allegheny & Butler & Lawrence \\
\hline
89-93 & $(x, \log^{2}(x))$ & $\log(x)$ & - & $(x, \log^{2}(x))$ \\
94-98 & $(x, \log(x), \log^{2}(x))$ & $(x, \log^{2}(x))$ & $(x, \log(x), \log^{2}(x))$ & $(x, \log(x), \log^{2}(x))$ \\
99-03 & $(x, \log^{2}(x))$ & $(x, \log(x), \log^{2}(x))$ & $(x, \log^{2}(x))$ & $(x, \log(x))$ \\
04-08 & $(x, \log^{2}(x))$ & $(x, \log^{2}(x))$ & - & $(x, \log^{2}(x))$ \\
09-13 & $(x, \log(x), \log^{2}(x))$ & $(x, \log(x), \log^{2}(x))$ & - & $(x, \log(x), \log^{2}(x))$ \\
14-17 & $(x, \log(x), \log^{2}(x))$ & $(x, \log^{2}(x))$ & $(x, \log^{2}(x))$ & $(x, \log^{2}(x))$ \\
\hline 
\end{tabular}
}
\label{Refined Tilts}
\end{table}

From the plots (not shown) of $\tilde{G}$ and $\hat{G}$ from all six periods,
it is observed that the empirical CDF's $\hat{G}$ are very 
close to each other and the estimated CDF's $\tilde{G}$ 
are even more similar. Therefore, 
it is sensible to combine the entire data from all the counties and all the periods 
to get an improved estimate of the reference $G$, and hence improved estimates of the
tail probabilities $1-G(T)$ for different thresholds $T$.

\begin{figure}[htbp]
\begin{center}
\includegraphics[width=3in]{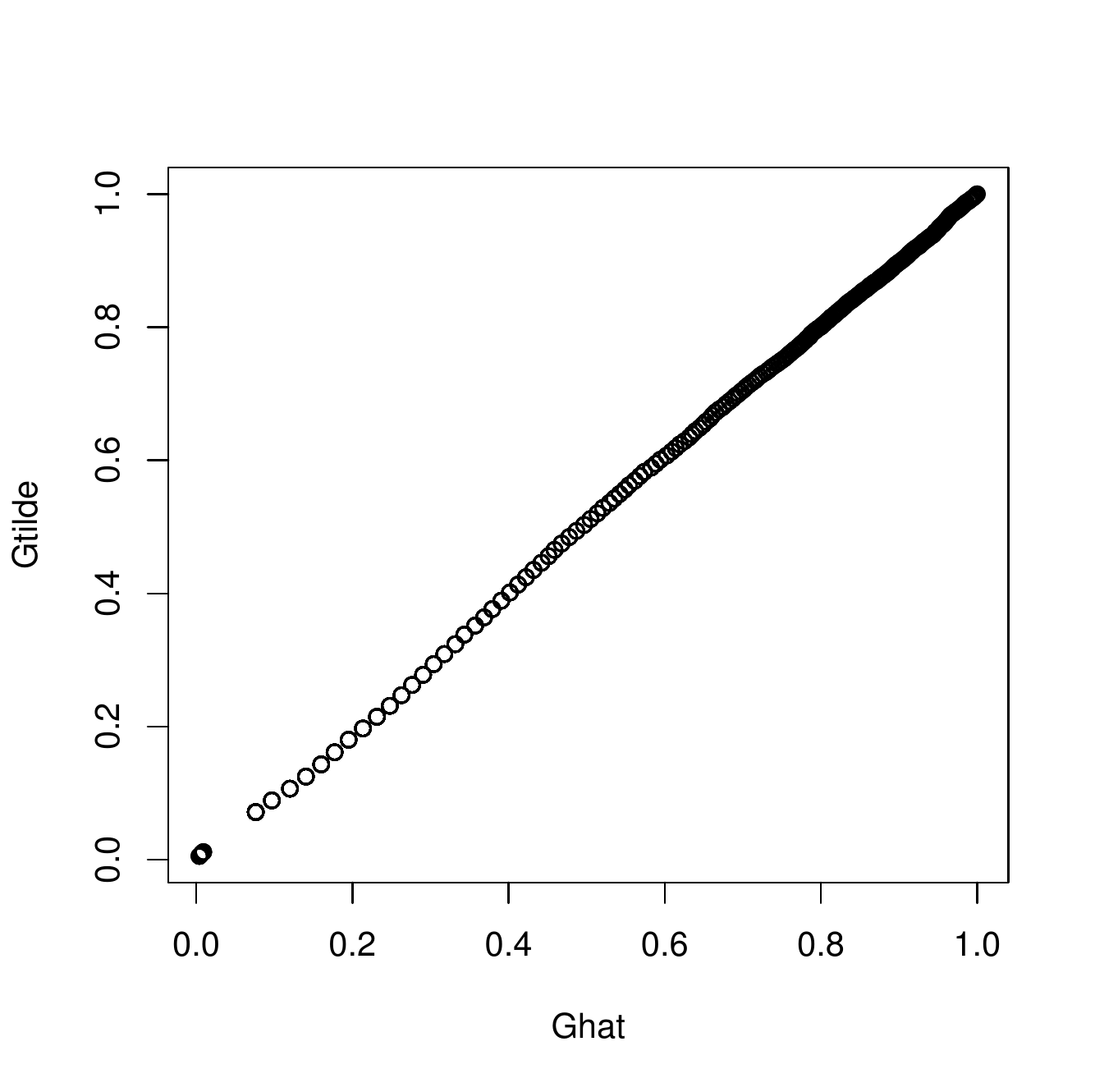}
\caption{$\tilde{G}$ versus $\hat{G}$}
\label{tildeG vs hatG}
\end{center}
\end{figure}

Thus, we use the combined data from all time periods and all five counties. We used the following variable tilt functions, suggested by first running pairwise fusions,
$\bm{h}_{1}(x)=\bm{h}_{4}(x)=(x, \log(x), \log^{2}(x))^T$ and 
$\bm{h}_{2}(x)=\bm{h}_{3}(x)=(x, \log^{2}(x))^T$. From the following 
goodness-of-fit plot \ref{tildeG vs hatG}, 
it is observed that the pairs $(\hat{G},\tilde{G})$ lie essentially on a 
$45^{\circ}$-line, pointing to a satisfactory DRM.

Finally, 
in Table \ref{Confidence Intervals}, 
we display Beaver threshold probability estimates for different thresholds $T$
and their corresponding confidence intervals obtained from 
(\ref{Variance of hat(G)}) in the Appendix. Additionally, we also display the empirical estimates and the corresponding confidence intervals to make a comparison with the density ratio estimates. 
{\em It is readily seen that the confidence intervals obtained from
the fused data using $\tilde{G}(T)$ obtained with variable tilts are shorter.}
Similar analysis can be repeated for any 
county and its immediate neighbors.

In general, an advantage of using $\tilde{G}$ over $\hat{G}$ is that
$\tilde{G}$ is smoother and has a greater support and hence can give
more information about the tail of the reference distribution.

\begin{table}[H]
\centering
\captionsetup{margin=10pt,font=small,labelfont=bf}
\caption{Beaver threshold probability estimates and 
95\% confidence intervals for threshold $T=5,10,25,50,100,150,200$. The intervals obtained from
the fused data using $\tilde{G}(T)$ are shorter.}
\scalebox{0.9}{
\begin{tabular}{ccc}
\hline
 $T$ & $1-\tilde{G}(T)$ & 95\% CI \\
 \hline
 5 & $3.991\times 10^{-1}$ & $(3.896\times 10^{-1}, 4.085\times 10^{-1})$ \\
 10 & $2.175\times 10^{-1}$ & $(2.093\times 10^{-1}, 2.256\times 10^{-1})$ \\
 25 & $7.001\times 10^{-2}$ & $(6.530\times 10^{-2}, 7.471\times 10^{-2})$ \\
 50 & $1.920\times 10^{-2}$ & $(1.666\times 10^{-2}, 2.175\times 10^{-2})$ \\
 100 & $2.697\times 10^{-3}$ & $(1.854\times 10^{-3}, 3.540\times 10^{-3})$\\
 150 & $5.727\times 10^{-4}$ & $(2.773\times 10^{-4}, 8.681\times 10^{-4})$\\
 200 & $1.263\times 10^{-4}$ & $(4.889\times 10^{-5}, 2.037\times 10^{-4})$\\
 \hline 
  $T$ & $1-\hat{G}(T)$ & 95\% CI \\
 \hline
 5 & $4.057\times 10^{-1}$ & $(3.945\times 10^{-1}, 4.168\times 10^{-1})$ \\
 10 & $2.183\times 10^{-1}$ & $(2.089\times 10^{-1}, 2.277\times 10^{-1})$ \\
 25 & $6.667\times 10^{-2}$ & $(6.099\times 10^{-2}, 7.234\times 10^{-2})$ \\
 50 & $1.980\times 10^{-2}$ & $(1.663\times 10^{-2}, 2.297\times 10^{-2})$ \\
 100 & $2.559\times 10^{-3}$ & $(1.410\times 10^{-3}, 3.708\times 10^{-3})$\\
 150 & $6.734\times 10^{-4}$ & $(8.335\times 10^{-5}, 1.263\times 10^{-3})$\\
 200 & $2.693\times 10^{-4}$ & $(-1.039\times 10^{-4}, 6.426\times 10^{-4})$\\
 \hline
\end{tabular}
}
\label{Confidence Intervals}
\end{table}

\section{Summary}

The density ratio model and its theoretical underpinnings have been discussed in the context of
case-control studies in quite a few works including 
\cite{FKQS},\cite{SDF},\cite{Lu2007},\cite{Qin2017},\cite{QinZhang97}, using {\em fixed tilts}.
Here we have applied an extension, which uses  {\em variable tilts}, in the estimation of 
threshold probabilities of radon concentration, fusing or combining data from several
Pennsylvania counties. We have focused on Beaver county and its 
four immediate neighbors, and relatively high thresholds much beyond the cutoff point of
4 picocurie per liter (4 pCi/L) \cite{EPA2003}. The results in Table \ref{Confidence Intervals} 
using a 
data fusion method are useful for 
public health policy and real estate property appraisal.

\clearpage
\begin{appendices}

\section{Appendix: Estimation and Asymptotic Results}
\label{Appendix}

The estimation and asymptotic results of the two-sample DRM with the tilt function $h(x)=x$ 
were established in \cite{QinZhang97} and the analysis of the two-sample case with a general tilt can be found in \cite{Zhang2000b}. An extension to the multi-sample case with an
unvaried or uniform tilt is proposed in \cite{FKQS} and a more detailed discussion can be found in \cite{SDF},\cite{Lu2007}.
The results in this section are thoroughly examined in \cite{Lu2007} under the assumption that the tilt functions are the same for each pair of samples:
(Beaver, Washington), (Beaver, Allegheny), (Beaver, Butler), (Beaver, Lawrence).
We extend the results in \cite{Lu2007} to account for variable tilts.

\subsection{Estimation}
Let $\bm{X}_{0}, \dots, \bm{X}_{m}$ be the samples with size $n_0, \dots, n_m$ respectively where $\bm{X}_{0}$ denotes the reference sample. Combine the data and denote it as $\bm{t}=(\bm{X}^{T}_{0}, \dots, \bm{X}^{T}_{m})^T$ with size $n=\sum_{k=0}^{m}n_k$. Suppose $\bm{X}_{k}\sim g_{k}$ for $k=0, \dots, m$ and they have the density ratio structure as (1). Denote the reference CDF
 as $G$ and let $w_k(\cdot)=\exp{(\alpha_k+\bm{\beta}^{T}_{k}\bm{h}_{k}(\cdot))}$ for $k=1, \dots, m$. We assume that
\begin{equation}
\int\left|\bm{h}_{i}(t)\right|w_{k}(t)dG(t)<\infty \phantom{aaaaa} \int\left|\bm{h}_{i}(t)\bm{h}^{T}_{j}(t)\right|w_{k}(t)dG(t)<\infty 
\end{equation}
$\forall i, j=1, \dots, m$ and $k=0, \dots, m$ where $w_0(\cdot)\equiv1$.

Let $\bm{\alpha}=(\alpha_1, \dots, \alpha_m)^T$, $\bm{\beta}=(\bm{\beta}^{T}_{1}, \dots, \bm{\beta}^{T}_{m})^T$ and $\bm{\theta}=(\bm{\alpha}^T, \bm{\beta}^T)^T$. Let $p_i=dG(t_i)$ for $i=1, \dots, n$. We wish to obtain the estimators $\tilde{\bm{\alpha}}$, $\tilde{\bm{\beta}}$ and $\tilde{p}_i$'s by maximizing
\begin{equation}
L(\bm{\theta}, G)=\prod_{i=1}^{n}p_i\prod_{k=1}^{m}\prod_{j=1}^{n_k}w_{k}(X_{kj})
\end{equation}
subject to
\begin{equation}
\sum_{i=1}^{n}p_i=1 \phantom{aaa}\sum_{i=1}^{n}p_{i}[w_{k}(t_i)-1]=0 \phantom{aaa} k=1, \dots, m.
\end{equation}
Then the objective function with Lagrange multipliers $\lambda_0, \dots, \lambda_m$ becomes
\begin{equation}
\log{L(\bm{\theta}, G)}-\lambda_{0}(1-\sum_{i=1}^{n}p_i)-\dots-\lambda_{m}\sum_{i=1}^{n}p_{i}[w_{m}(t_i)-1].
\end{equation}
Differentiate the objective function with respect to $p_i$ for $i=1, \dots, n$, we obtain
\begin{equation}
1+\lambda_{0}p_{i}-\lambda_{1}p_{i}[w_{1}(t_i)-1]-\dots-\lambda_{m}[w_{m}(t_i)-1]=0.
\label{est1}
\end{equation}
Sum over $i$, we have $\lambda_0=-n$. Replace $\lambda_0$ with $-n$ in (\ref{est1}), we have
\begin{equation}
p_i=\frac{1}{n+\lambda_{1}p_{i}[w_{1}(t_i)-1]+\dots+\lambda_{m}[w_{m}(t_i)-1]}
\label{est2}
\end{equation}
for $i=1, \dots, n$.
Substitute (\ref{est2}) into $\log{L(\bm{\theta}, G)}$, we have
\begin{equation}
\begin{split}
\log{L(\bm{\theta}, G)}=&-\sum_{i=1}^{n}\log{(n+\lambda_{1}p_{i}[w_{1}(t_i)-1]+\dots+\lambda_{m}[w_{m}(t_i)-1])}\\
&+\sum_{k=1}^{m}(n_{k}\alpha_{k}+\sum_{j=1}^{n_k}\bm{\beta}^{T}_{k}\bm{h}_{k}(X_{kj})).
\end{split}
\label{est3}
\end{equation}
Differentiate $\log{L(\bm{\theta}, G)}$ with respect to $\alpha_k$ for $k=1, \dots, m$,
\begin{equation}
\begin{split}
\frac{\partial \log{L(\bm{\theta}, G)}}{\partial \alpha_k}&=-\sum_{i=1}^{n}\frac{\lambda_{k}w_{k}(t_i)}{n+\lambda_{1}p_{i}[w_{1}(t_i)-1]+\dots+\lambda_{m}[w_{m}(t_i)-1]}+n_k \\
&=-\sum_{i=1}^{n}\lambda_{k}p_{i}w_{k}(t_i)+n_k \\
&=-\lambda_{k}+n_k. 
\end{split}
\end{equation}
Therefore, we obtain $\lambda_k=n_k$ by equating $\frac{\partial \log{L(\bm{\theta}, G)}}{\partial \alpha_k}=0$ for $k=1, \dots, m$. Substitute $\lambda_k$'s into (\ref{est2}) and denote $\rho_{k} \equiv \frac{n_k}{n_0}$ for $k=0, \dots, m$, we have
\begin{equation}
p_i=\frac{1}{n_{0}\sum_{k=0}^{m}\rho_{k}w_{k}(t_i)}
\end{equation}
Substitute $p_i$'s into (\ref{est3}), $\log{L(\bm{\theta}, G)}$ reduces to a function of $\bm{\theta}$ only.
\begin{equation}
\begin{split}
l(\bm{\theta})\equiv \log{L(\bm{\theta}, G)}=&-\sum_{i=1}^{n}\log{(n_{0}\sum_{k=0}^{m}\rho_{k}w_{k}(t_i))}\\
&+\sum_{k=1}^{m}(n_{k}\alpha_{k}+\sum_{j=1}^{n_k}\bm{\beta}^{T}_{k}\bm{h}_{k}(X_{kj})).
\end{split}
\end{equation}
Then we obtain the estimators $\tilde{\bm{\theta}}=(\tilde{\bm{\alpha}}^{T}, \tilde{\bm{\beta}}^{T})^T$ via solving the system of equations
\begin{equation}
\begin{split}
&\frac{\partial l(\bm{\theta})}{\partial \alpha_k}=-\sum_{i=1}^{n}\frac{\rho_{k}w_{k}(t_i)}{\sum_{k=0}^{m}\rho_{k}w_{k}(t_i)}+n_{k}=0 \\
&\frac{\partial l(\bm{\theta})}{\partial \bm{\beta}_{k}}=-\sum_{i=1}^{n}\frac{\rho_{k}w_{k}(t_i)\bm{h}_{k}(t_{i})}{\sum_{k=0}^{m}\rho_{k}w_{k}(t_i)}+\sum_{j=1}^{n_k}\bm{h}_{k}(X_{kj})=\bm{0}.
\end{split}
\end{equation}
Then obtain the estimators $\tilde{p}_i$ for $i=1, \dots, n$ such that
\begin{equation}
\tilde{p}_i=\frac{1}{n_{0}\sum_{k=0}^{m}\rho_{k}\tilde{w}_{k}(t_i)}
\end{equation}
where $\tilde{w}_{k}(\cdot)=\exp{(\tilde{\alpha}_{k}+\tilde{\bm{\beta}}^{T}_{k}\bm{h}_{k}(\cdot))}$ for $k=1, \dots, m$ and $\tilde{w}_{0}(\cdot)\equiv1$. Thus, the estimator 
of the reference CDF $G$ is
\begin{equation}
\tilde{G}(t)=\sum_{i=1}^{n}\tilde{p}_{i}I(t_{i}\leq t).
\end{equation}
\subsection{Asymptotic Distribution of $\tilde{\bm{\theta}}$}
In this section, we shall establish the strong consistency and the asymptotic normality of $\tilde{\bm{\theta}}$. Assume that $l(\bm{\theta})$ is  concave and second-order differentiable, the density ratio model is non-degenerate and $\rho_k$'s are fixed as $n\to 0$.  Let $\bm{\theta}_0$ be the true parameter vector. We shall start with obtaining the expectations of the first order derivatives for $k=1, \dots, m$
\begin{equation}
\begin{split}
\mathrm{E}\bigg(\frac{\partial l}{\partial \alpha_k}\bigg)&=-\sum_{i=1}^{n}\mathrm{E}\bigg(\frac{\rho_{k}w_{k}(t_i)}{\sum_{k=0}^{m}\rho_{k}w_{k}(t_i)}\bigg)+n_k \\
&=-\sum_{u=0}^{m}n_{u}\int\frac{\rho_{k}w_{k}(t)}{\sum_{k=0}^{m}\rho_{k}w_{k}(t)}w_{u}(t)dG(t)+n_k \\
&=0
\end{split}
\end{equation}
\begin{equation}
\begin{split}
\mathrm{E}\bigg(\frac{\partial l}{\partial \bm{\beta}_k}\bigg)&=-\sum_{i=1}^{n}\mathrm{E}\bigg(\frac{\rho_{k}w_{k}(t_i)\bm{h}_{k}(t_{i})}{\sum_{k=0}^{m}\rho_{k}w_{k}(t_i)}\bigg)+\sum_{j=1}^{n_k}E\bm{h}_{k}(X_{kj}) \\
&=-\sum_{u=0}^{m}n_{u}\int\frac{\rho_{k}w_{k}(t)\bm{h}_{k}(t)}{\sum_{j=0}^{m}\rho_{j}w_{j}(t)}w_{u}(t)dG(t) \\
&\phantom{aa}+n_{k}\int w_{k}(t)\bm{h}_{k}(t)dG(t) \\
&=\bm{0}.
\end{split}
\end{equation}
By the strong law of large number, for $k=1, \dots, m$,
\begin{equation}
\begin{split}
&\frac{1}{n}\frac{\partial l}{\partial \alpha_k}=-\frac{n_0}{n}\sum_{u=0}^{m}\rho_{u}\frac{1}{n_u}\sum_{j=1}^{n_u}\frac{\rho_{k}w_{k}(X_{uj})}{\sum_{k=0}^{m}\rho_{k}w_{k}(X_{uj})}+\frac{\rho_k}{\sum_{k=0}^{m}\rho_k} \\
&\phantom{aaaaa}\stackrel{a.s.}{\to}-\frac{1}{\sum_{k=0}^{m}\rho_k}\sum_{u=1}^{m}\rho_{u}\int\frac{\rho_{k}w_{k}(t)}{\sum_{k=0}^{m}\rho_{k}w_{k}(t)}w_{u}(t)dG(t)+\frac{\rho_k}{\sum_{k=0}^{m}\rho_k} \\
&\phantom{aaaaa}=0\\
&\frac{1}{n}\frac{\partial l}{\partial \bm{\beta}_k}\stackrel{a.s.}{\to}-\frac{1}{\sum_{k=0}^{m}\rho_k}\sum_{u=1}^{m}\rho_{u}\int\frac{\rho_{k}w_{k}(t)\bm{h}_{k}(t)}{\sum_{k=0}^{m}\rho_{k}w_{k}(t)}w_{u}(t)dG(t) \\
&\phantom{aaaaaaaa}+\frac{\rho_k}{\sum_{k=0}^{m}\rho_k}\int w_{k}(t)\bm{h}_{k}(t)dG(t) \\
&\phantom{aaaaa}=0.
\end{split}
\end{equation}
Denote 
\begin{equation}
\begin{split}
&l_{1}(\bm{\theta})=\sum_{i=1}^{n}\log\sum_{k=0}^{m}\rho_{k}w_{k}(t_i) \\
&l_{2}(\bm{\theta})=\sum_{k=1}^{m}(n_{k}\alpha_{k}+\sum_{j=1}^{n_k}\bm{\beta}^{T}_{k}\bm{h}_{k}(X_{kj})),
\end{split}
\end{equation}
and then we have $l=-n\log{n_0}-l_{1}+l_2$. Since $\frac{1}{n}\frac{\partial l}{\partial \bm{\theta}}\stackrel{a.s.}{\to}\bm{0}$, then
\begin{equation}
\frac{1}{n}\left|\frac{\partial l_1}{\partial \bm{\theta}}-\frac{\partial l_2}{\partial \bm{\theta}}\right|\stackrel{a.s.}{\to}\bm{0}.
\label{ASL12}
\end{equation}
For $l_1$, we have
\begin{equation}
\frac{\partial^{2} l_1}{\partial\bm{\theta}\partial\bm{\theta}^T}=-\frac{\partial^{2} l}{\partial\bm{\theta}\partial\bm{\theta}^T}
\end{equation}
which indicates that $\frac{\partial^{2} l_1}{\partial\bm{\theta}\partial\bm{\theta}^T}$ is positive definite since $l$ is concave and the model is non-degenerate. 

For $l_2$,
\begin{equation}
\frac{\partial l_2}{\partial\bm{\theta}}=(n_{1}, \dots, n_{m}, \sum_{j=1}^{n_1}\bm{h}^{T}_{1}(X_{1j}), \dots, \sum_{j=1}^{n_m}\bm{h}^{T}_{m}(X_{mj}))^T
\end{equation}
so that $\frac{\partial l_2}{\partial\bm{\theta}}$ is independent of $\bm{\theta}$ and 
\begin{equation}
l_2=\bm{\theta}^{T}\frac{\partial l_2}{\partial\bm{\theta}}.
\label{linear}
\end{equation}

To prove the strong consistency, it is sufficient to show that the maximum of $l$ is not obtained on the boundary of any closed sphere of $\bm{\theta}_0$ almost surely. We show for sufficiently large $n$, $\forall \varepsilon, l(\bm{\theta}_0)>l(\bm{\theta}^{\ast})$ almost surely for all $\bm{\theta}^{\ast}\in \bar{B}(\bm{\theta}_0, \varepsilon)$.

Expand $\frac{1}{n}l_1$ at $\bm{\theta}_0$,
\begin{equation}
\frac{1}{n}l_{1}(\bm{\theta}^{\ast})=\frac{1}{n}l_{1}(\bm{\theta}_0)+\frac{1}{n}(\bm{\theta}^{\ast}-\bm{\theta}_0)^{T}\frac{1}{n}\frac{\partial l_{1}(\bm{\theta}_0)}{\partial \bm{\theta}}+\frac{1}{2n}(\bm{\theta}^{\ast}-\bm{\theta}_0)^{T}\frac{\partial^{2} l_{1}(\bm{\theta}^{\ast\ast})}{\partial\bm{\theta}\partial\bm{\theta}^T}(\bm{\theta}^{\ast}-\bm{\theta}_0)
\end{equation}
where $\bm{\theta}^{\ast\ast}$ is between $\bm{\theta}_0$ and $\bm{\theta}^{\ast}$. Since that $\frac{\partial^{2} l_1}{\partial\bm{\theta}\partial\bm{\theta}^T}$ is positive definite, then we have
\begin{equation}
\frac{1}{n}l_{1}(\bm{\theta}^{\ast})-\frac{1}{n}l_{1}(\bm{\theta}_0)-\frac{1}{n}(\bm{\theta}^{\ast}-\bm{\theta}_0)^{T}\frac{1}{n}\frac{\partial l_{1}(\bm{\theta}_0)}{\partial \bm{\theta}}>0.
\end{equation}
By (\ref{ASL12}), for sufficiently large $n$, 
\begin{equation}
\frac{1}{n}l_{1}(\bm{\theta}^{\ast})-\frac{1}{n}l_{1}(\bm{\theta}_0)-\frac{1}{n}(\bm{\theta}^{\ast}-\bm{\theta}_0)^{T}\frac{1}{n}\frac{\partial l_{2}(\bm{\theta}_0)}{\partial \bm{\theta}}>0.
\end{equation}
Since that $\frac{\partial l_2}{\partial\bm{\theta}}$ is independent of $\bm{\theta}$ and (\ref{linear}), then
\begin{equation}
\begin{split}
\frac{1}{n}l_{1}(\bm{\theta}^{\ast})-\frac{1}{n}l_{1}(\bm{\theta}_0)-\frac{1}{n}l_{2}(\bm{\theta}^{\ast})+\frac{1}{n}l_{2}(\bm{\theta}_0)&>0 \\
\frac{1}{n}l_{1}(\bm{\theta}_0)-\frac{1}{n}l_{1}(\bm{\theta}^{\ast})&>0
\end{split}
\end{equation}
almost surely.
Therefore, the strong consistency of the estimator $\tilde{\bm{\theta}}$ is established.

Now we wish to obtain the limit of $-\frac{1}{n}\frac{\partial^{2} l}{\partial\bm{\theta}\partial\bm{\theta}^T}$. By the strong law of large number, for $k,k^{\prime}=1, \dots, m$ and $k\neq k^{\prime}$, as $n\to\infty$,
\begin{equation}
\begin{split}
&-\frac{1}{n}\frac{\partial^{2} l}{\partial \alpha^{2}_{k}}\stackrel{a.s.}{\to}\frac{\rho_k}{\sum_{j=0}^{m}\rho_j}-\frac{\rho^{2}_{k}}{\sum_{j=0}^{m}\rho_j}\int\frac{w^{2}_{k}(t)}{\sum_{j=0}^{m}\rho_{j}w_{j}(t)}dG(t)\\
&-\frac{1}{n}\frac{\partial^{2} l}{\partial \alpha_{k}\partial \alpha_{k^{\prime}}}\stackrel{a.s.}{\to}-\frac{\rho_{k}\rho_{k^{\prime}}}{\sum_{j=0}^{m}\rho_j}\int\frac{w_{k}(t)w_{k^{\prime}}(t)}{\sum_{j=0}^{m}\rho_{j}w_{j}(t)}dG(t)\\
&-\frac{1}{n}\frac{\partial^{2} l}{\partial \alpha_{k}\partial \bm{\beta}^{T}_{k}}\stackrel{a.s.}{\to}\frac{\rho_k}{\sum_{j=0}^{m}\rho_j}\int w_{k}(t)\bm{h}^{T}_{k}(t)dG(t) \\
&\phantom{aaaaaaaaaaaaaa}-\frac{\rho^{2}_{k}}{\sum_{j=0}^{m}\rho_j}\int\frac{w^{2}_{k}(t)\bm{h}^{T}_{k}(t)}{\sum_{j=0}^{m}\rho_{j}w_{j}(t)}dG(t)\\
&-\frac{1}{n}\frac{\partial^{2} l}{\partial \alpha_{k}\partial \bm{\beta}^{T}_{k^{\prime}}}\stackrel{a.s.}{\to}-\frac{\rho_{k}\rho_{k^{\prime}}}{\sum_{j=0}^{m}\rho_j}\int\frac{w_{k}(t)w_{k^{\prime}}(t)\bm{h}^{T}_{k^{\prime}}(t)}{\sum_{j=0}^{m}\rho_{j}w_{j}(t)}dG(t) \\
&-\frac{1}{n}\frac{\partial^{2} l}{\partial \bm{\beta}_{k}\partial \bm{\beta}^{T}_{k}}\stackrel{a.s.}{\to}\frac{\rho_k}{\sum_{j=0}^{m}\rho_j}\int w_{k}(t)\bm{h}_{k}(t)\bm{h}^{T}_{k}(t)dG(t) \\
&\phantom{aaaaaaaaaaaaaa}-\frac{\rho^{2}_{k}}{\sum_{j=0}^{m}\rho_j}\int\frac{w^{2}_{k}(t)\bm{h}_{k}(t)\bm{h}^{T}_{k}(t)}{\sum_{j=0}^{m}\rho_{j}w_{j}(t)}dG(t)\\
&-\frac{1}{n}\frac{\partial^{2} l}{\partial \bm{\beta}_{k}\partial \bm{\beta}^{T}_{k^{\prime}}}\stackrel{a.s.}{\to}-\frac{\rho_{k}\rho_{k^{\prime}}}{\sum_{j=0}^{m}\rho_j}\int\frac{w_{k}(t)w_{k^{\prime}}(t)\bm{h}_{k}(t)\bm{h}^{T}_{k^{\prime}}(t)}{\sum_{j=0}^{m}\rho_{j}w_{j}(t)}dG(t).
\end{split}
\end{equation}
Define the following quantities for $k, k^{\prime}=1, \dots, m$
\begin{equation}
\begin{split}
&A_{kk^{\prime}}=\int\frac{w_{k}(t)w_{k^{\prime}}(t)}{\sum_{j=0}^{m}\rho_{j}w_{j}(t)}dG(t) \\
&\bm{B}_{kk^{\prime}}=\int\frac{w_{k}(t)w_{k^{\prime}}(t)\bm{h}^{T}_{k^{\prime}}(t)}{\sum_{j=0}^{m}\rho_{j}w_{j}(t)}dG(t) \\
&\bm{C}_{kk^{\prime}}=\int\frac{w_{k}(t)w_{k^{\prime}}(t)\bm{h}_{k}(t)\bm{h}^{T}_{k^{\prime}}(t)}{\sum_{j=0}^{m}\rho_{j}w_{j}(t)}dG(t) \\
&\bm{E}_{k}=\int w_{k}(t)\bm{h}_{k}(t)dG(t) \\
&\bar{\bm{E}}_{k}=\int w_{k}(t)\bm{h}_{k}(t)\bm{h}^{T}_{k}(t)dG(t) \\
&\bm{V}_{k}=\bar{\bm{E}}_{k}-\bm{E}_{k}\bm{E}^{T}_{k} \\
&\bm{A}=(A_{ij})_{m\times m} \phantom{aaa}\bm{B}=(\bm{B}_{ij})_{m\times r} \phantom{aaa}\bm{C}=(\bm{C}_{ij})_{r\times r}\\
&\bm{\rho}=\left[ 
\begin{matrix}
\rho_{1} & \cdots & 0 \\
\vdots & \ddots & \vdots \\
0 & \cdots & \rho_{m}
\end{matrix}
\right]_{m\times m} \phantom{aaa} \bar{\bm{\rho}}=\left[ 
\begin{matrix}
\rho_{1}\bm{I}_{r_1} & \cdots & \bm{0} \\
\vdots & \ddots & \vdots \\
\bm{0} & \cdots & \rho_{m}\bm{I}_{r_m}
\end{matrix}
\right]_{r\times r} \\
&\bm{E}=\left[ 
\begin{matrix}
\bm{E}_{1} & \cdots & \bm{0} \\
\vdots & \ddots & \vdots \\
\bm{0} & \cdots & \bm{E}_{m}
\end{matrix}
\right]_{r\times m} \phantom{aaa} \bar{\bm{E}}=\left[ 
\begin{matrix}
\bar{\bm{E}}_{1} & \cdots & \bm{0} \\
\vdots & \ddots & \vdots \\
\bm{0} & \cdots & \bar{\bm{E}}_{m}
\end{matrix}
\right]_{r\times r} \\
&\bm{V}=\left[ 
\begin{matrix}
\bm{V}_{1} & \cdots & \bm{0} \\
\vdots & \ddots & \vdots \\
\bm{0} & \cdots & \bm{V}_{m}
\end{matrix}
\right]_{r\times r}
\end{split}
\end{equation}
where $r_k$ is the dimension of $\bm{h}_k$ and $r=\sum_{k=1}^{m}r_{k}$. Therefore, 
\begin{equation}
-\frac{1}{n}\frac{\partial^{2}l}{\partial \bm{\theta} \partial \bm{\theta}^{T}}\stackrel{a.s.}{\to}\bm{S}=\frac{1}{\sum_{k=0}^{m}\rho_k}\left[
\begin{matrix}
\bm{S}_{11} & \bm{S}_{12} \\
\bm{S}^{T}_{12} & \bm{S}_{22}
\end{matrix}
\right]
\end{equation}
where
\begin{equation}
\begin{split}
&\bm{S}_{11}=\bm{\rho}-\bm{\rho}\bm{A}\bm{\rho} \\
&\bm{S}_{12}=\bm{\rho}\bm{E}^{T}-\bm{\rho}\bm{B}\bar{\bm{\rho}} \\
&\bm{S}_{22}=\bar{\bm{\rho}}\bar{\bm{E}}-\bar{\bm{\rho}}\bm{C}\bar{\bm{\rho}}.
\end{split}
\end{equation}
Next, obtain the expression of  $\mathrm{Var}(\frac{1}{\sqrt{n}}\frac{\partial l}{\partial \bm{\theta}})$. For $k, k^{\prime}=1, \dots, m$ and $k\neq k^{\prime}$,
\begin{equation}
\begin{split}
&\mathrm{Var}\bigg(\frac{1}{\sqrt{n}}\frac{\partial l}{\partial \alpha_k}\bigg)=\frac{\rho^{2}_{k}}{\sum_{j=0}^{m}\rho_j}(A_{kk}-\sum_{j=1}^{m}\rho_{j}A^{2}_{kj}-(1-\sum_{j=1}^{m}\rho_{j}A_{kj})^{2}) \\
&\mathrm{Cov}\bigg(\frac{1}{\sqrt{n}}\frac{\partial l}{\partial \alpha_{k}}, \frac{1}{\sqrt{n}}\frac{\partial l}{\partial \alpha_{k^{\prime}}}\bigg)=\frac{\rho_{k}\rho_{k^{\prime}}}{\sum_{j=0}^{m}\rho_j}(A_{kk^{\prime}}-\sum_{j=1}^{m}\rho_{j}A_{kj}A_{k^{\prime}j} \\
&\phantom{aaaaaaaaaaaaaaaaaaaaaaa}-(1-\sum_{j=1}^{m}\rho_{j}A_{kj})(1-\sum_{j=1}^{m}\rho_{j}A_{k^{\prime}j})) \\
&\mathrm{Cov}\bigg(\frac{1}{\sqrt{n}}\frac{\partial l}{\partial \alpha_{k}}, \frac{1}{\sqrt{n}}\frac{\partial l}{\partial \bm{\beta}_{k}}\bigg)=\frac{\rho^{2}_{k}}{\sum_{j=0}^{m}\rho_j}(A_{kk}\bm{E}_{k}-\sum_{j=1}^{m}\rho_{j}A_{kj}\bm{B}^{T}_{jk}\\
&\phantom{aaaaaaaaaaaaaaaaaaaaaaa}-(1-\sum_{j=1}^{m}\rho_{j}A_{kj})(\bm{E}_{k}-\sum_{j=1}^{m}\rho_{j}\bm{B}^{T}_{jk})) \\
&\mathrm{Cov}\bigg(\frac{1}{\sqrt{n}}\frac{\partial l}{\partial \alpha_{k}}, \frac{1}{\sqrt{n}}\frac{\partial l}{\partial \bm{\beta}_{k^{\prime}}}\bigg)=\frac{\rho_{k}\rho_{k^{\prime}}}{\sum_{j=0}^{m}\rho_j}(A_{kk^{\prime}}\bm{E}_{k^{\prime}}-\sum_{j=1}^{m}\rho_{j}A_{kj}\bm{B}^{T}_{jk^{\prime}}\\
&\phantom{aaaaaaaaaaaaaaaaaaaaaaa}-(1-\sum_{j=1}^{m}\rho_{j}A_{kj})(\bm{E}_{k^{\prime}}-\sum_{j=1}^{m}\rho_{j}\bm{B}^{T}_{jk^{\prime}})) \\
&\mathrm{Var}\bigg(\frac{1}{\sqrt{n}}\frac{\partial l}{\partial \bm{\beta}_k}\bigg)=\frac{\rho^{2}_{k}}{\sum_{j=0}^{m}\rho_j}(-\bm{C}_{kk}-\sum_{j=1}^{m}\rho_{j}\bm{B}^{T}_{jk}\bm{B}_{jk}-(\bm{E}_{k}-\sum_{j=1}^{m}\rho_{j}\bm{B}^{T}_{jk})\\
&\phantom{aaaaaaaaaaaaaaa}(\bm{E}^{T}_{k}-\sum_{j=1}^{m}\rho_{j}\bm{B}_{jk})+2\bm{B}^{T}_{kk}\bm{E}^{T}_{k})+\frac{\rho_{k}}{\sum_{j=0}^{m}\rho_j}\bm{V}_{k}\\
&\mathrm{Cov}\bigg(\frac{1}{\sqrt{n}}\frac{\partial l}{\partial \bm{\beta}_{k}}, \frac{1}{\sqrt{n}}\frac{\partial l}{\partial \bm{\beta}_{k^{\prime}}}\bigg)=\frac{\rho_{k}\rho_{k^{\prime}}}{\sum_{j=0}^{m}\rho_j}(-\bm{C}_{kk^{\prime}}-\sum_{j=1}^{m}\rho_{j}\bm{B}^{T}_{jk}\bm{B}_{jk^{\prime}}\\
&\phantom{aaaaaaaaaaaaaaaaaaaaaaa}-(\bm{E}_{k}-\sum_{j=1}^{m}\rho_{j}\bm{B}^{T}_{jk})(\bm{E}^{T}_{k^{\prime}}-\sum_{j=1}^{m}\rho_{j}\bm{B}_{jk^{\prime}})\\
&\phantom{aaaaaaaaaaaaaaaaaaaaaaa}+\bm{B}^{T}_{kk^{\prime}}\bm{E}^{T}_{k}+\bm{B}^{T}_{kk^{\prime}}\bm{E}^{T}_{k^{\prime}}).
\end{split}
\end{equation}
Let $\Lambda\equiv\mathrm{Var}(\frac{1}{\sqrt{n}}\frac{\partial l}{\partial \bm{\theta}})$ and $\Lambda\equiv\frac{1}{\sum_{k=0}^{m}\rho_k}\left[
\begin{matrix}
\bm{\Lambda}_{11} & \bm{\Lambda}_{12} \\
\bm{\Lambda}^{T}_{12} & \bm{\Lambda}_{22}
\end{matrix}
\right]$, then we have
\begin{equation}
\begin{split}
&\bm{\Lambda}_{11}=\bm{\rho}(\bm{A}-\bm{A}\bm{\rho}\bm{A}-(\bm{I}_{m}-\bm{A}\bm{\rho})\bm{J}_{m}(\bm{I}_{m}-\bm{\rho}\bm{A}))\bm{\rho}\\
&\phantom{aaa}=\bm{S}_{11}-\bm{S}_{11}(\bm{J}_{m}+\bm{\rho}^{-1})\bm{S}_{11}\\
&\bm{\Lambda}_{12}=\bm{\rho}(\bm{A}\bm{E}^{T}-\bm{A}\bm{\rho}\bm{B}-(\bm{I}_{m}-\bm{A}\bm{\rho})\bm{J}_{m}(\bm{E}^{T}-\bm{\rho}\bm{B}))\bar{\bm{\rho}}\\
&\phantom{aaa}=\bm{S}_{12}-\bm{S}_{11}(\bm{J}_{m}+\bm{\rho}^{-1})\bm{S}_{12}\\
&\bm{\Lambda}_{22}=\bar{\bm{\rho}}(-\bm{C}-\bm{B}^{T}\bm{\rho}\bm{B}-(\bm{E}-\bm{B}^{T}\bm{\rho})\bm{J}_{m}(\bm{E}^{T}-\bm{\rho}\bm{B})+\bm{B}^{T}\bm{E}^{T}\\
&\phantom{aaaaa}+\bm{E}\bm{B})\bar{\bm{\rho}}+\bar{\bm{\rho}}(\bar{\bm{E}}-\bm{E}\bm{E}^{T})\\
&\phantom{aaa}=\bm{S}_{22}-\bm{S}_{21}(\bm{J}_{m}+\bm{\rho}^{-1})\bm{S}_{12}
\end{split}
\end{equation}
so that
\begin{equation}
\bm{\Sigma}\equiv\bm{S}^{-1}\bm{\Lambda}\bm{S}^{-1}=\bm{S}^{-1}-\sum_{j=0}^{m}\rho_{j}\left[
\begin{matrix}
\bm{J}_{m}+\bm{\rho}^{-1} & \bm{0} \\
\bm{0} & \bm{0}
\end{matrix}
\right].
\end{equation}

Expand $\frac{\partial l}{\partial \bm{\theta}}$ at $\bm{\theta}_{0}$ and plug in $\tilde{\bm{\theta}}$,
\begin{equation}
\begin{split}
\frac{\partial l(\tilde{\bm{\theta}})}{\partial \bm{\theta}}&=\frac{\partial l(\bm{\theta}_{0})}{\partial \bm{\theta}}+\frac{\partial^{2} l(\bm{\theta}_{0})}{\partial \bm{\theta}\partial\bm{\theta}^{T}}(\tilde{\bm{\theta}}-\bm{\theta}_{0})+o(||\tilde{\bm{\theta}}-\bm{\theta}_{0}||) \\
\sqrt{n}(\tilde{\bm{\theta}}-\bm{\theta}_{0})&=-\bigg(\frac{1}{n}\frac{\partial^{2} l(\bm{\theta}_{0})}{\partial \bm{\theta}\partial\bm{\theta}^{T}}\bigg)^{-1}\bigg(\frac{1}{\sqrt{n}}\frac{\partial l(\bm{\theta}_{0})}{\partial \bm{\theta}}\bigg)+o(1).
\end{split}
\end{equation}
By the central limit theorem,
\begin{equation}
\frac{1}{\sqrt{n}}\frac{\partial l(\bm{\theta}_{0})}{\partial \bm{\theta}}\stackrel{d}{\to}N(\bm{0}, \bm{\Lambda}).
\end{equation}
And since $-\frac{1}{n}\frac{\partial^{2} l(\bm{\theta}_{0})}{\partial \bm{\theta}\partial\bm{\theta}^{T}}\stackrel{a.s.}{\to}\bm{S}$, by the Slutsky's theorem,
\begin{equation}
\sqrt{n}(\tilde{\bm{\theta}}-\bm{\theta}_{0})\stackrel{d}{\to}N(\bm{0}, \bm{\Sigma}).
\end{equation}
Hence, the asymptotic normality of $\tilde{\bm{\theta}}$ is established.
\subsection{Asymptotic Behavior of $\tilde{G}$}
In this section, we wish to show that $\sqrt{n}(\tilde{G}-G)$ converges weakly to a zero-mean Gaussian process in $D[-\infty, \infty]$ with covariance function $R$. The proof in \cite{Lu2007} can be adopted with a slight modification. To prove the weak convergence of $\sqrt{n}(\tilde{G}-G)$, it is sufficient to show the weak convergence of $\sqrt{n}(\tilde{G}-\hat{G})$ due to the weak convergence of the empirical process $\sqrt{n}(\hat{G}-G)$. 

Let
\begin{equation}
\begin{split}
&A_{k}(t)=\int\frac{w_{k}(y)I[y\leq t]}{\sum_{j=0}^{m}\rho_{j}w_{j}(y)}dG(y) \\
&\bm{B}_{k}(t)=\int\frac{w_{k}(y)\bm{h}_{k}(y)I[y\leq t]}{\sum_{j=0}^{m}\rho_{j}w_{j}(y)}dG(y) \\
&\bar{\bm{A}}(t)=(A_{1}(t), \cdots, A_{m}(t))^{T} \\
&\bar{\bm{B}}(t)=(\bm{B}^{T}_{1}(t), \cdots, \bm{B}^{T}_{m}(t))^{T} \\
&H_{1}(t; \bm{\theta})=\frac{1}{n_0}\sum_{i=1}^{n}\frac{I[t_{i}\leq t]}{\sum_{j=0}^{m}\rho_{j}w_{j}(t_i)} \\
&H_{2}(t; \bm{\theta})=\frac{1}{n}(\bar{\bm{A}}^{T}(t)\bm{\rho}, \bar{\bm{B}}^{T}(t)\bar{\bm{\rho}})\bm{S}^{-1}\frac{\partial l(\bm{\theta})}{\partial \bm{\theta}}.
\end{split}
\end{equation}
Then we have
\begin{equation}
\begin{split}
&\mathrm{E}\bigg(\frac{\partial H_{1}(t; \bm{\theta}_{0})}{\partial\bm{\theta}}\bigg)=(\bar{\bm{A}}^{T}(t)\bm{\rho}, \bar{\bm{B}}^{T}\bar{\bm{\rho}})^{T},
\end{split}
\end{equation}
and by the strong law of large number, we can show
\begin{equation}
\frac{\partial H_{1}(t; \bm{\theta}_{0})}{\partial\bm{\theta}}\stackrel{a.s.}{\to}\mathrm{E}\bigg(\frac{\partial H_{1}(t; \bm{\theta}_{0})}{\partial\bm{\theta}}\bigg).
\end{equation} 
By the boundness of $\frac{w_{k}(y)}{\sum_{j=0}^{m}\rho_{j}w_{j}(y)}$ and $\frac{w_{k}(y)\bm{h}_{k}(y)}{\sum_{j=0}^{m}\rho_{j}w_{j}(y)}$, we have
\begin{equation}
\mathop{\mathrm{sup}}\limits_{-\infty<t<\infty}\bigg|\bigg|\frac{\partial H_{1}(t; \bm{\theta}_{0})}{\partial\bm{\theta}}-\mathrm{E}\bigg(\frac{\partial H_{1}(t; \bm{\theta}_{0})}{\partial\bm{\theta}}\bigg)\bigg|\bigg|\stackrel{a.s.}{\to}0.
\end{equation}

Expand $\tilde{G}$ at $\bm{\theta}_0$,
\begin{equation}
\begin{split}
&\tilde{G}(t)=H_{1}(t; \bm{\theta}_{0})+\bigg(\frac{\partial H_{1}(t; \bm{\theta}_{0})}{\partial\bm{\theta}}-\mathrm{E}\bigg(\frac{\partial H_{1}(t; \bm{\theta}_{0})}{\partial\bm{\theta}}\bigg)\bigg)^{T}(\tilde{\bm{\theta}}-\bm{\theta}_{0})\\
&\phantom{aaaaaa}-H_{2}(t; \bm{\theta}_{0})-(\bar{\bm{A}}^{T}(t)\bm{\rho}, \bar{\bm{B}}^{T}(t)\bar{\bm{\rho}})\bigg(\bm{\theta}_{0}-\frac{1}{n}\bm{S}^{-1}\frac{\partial l(\bm{\theta}_{0})}{\partial \bm{\theta}}\bigg)\\
&\phantom{aaaaaa}+o(||\tilde{\bm{\theta}}-\bm{\theta}_{0}||)\\
&\phantom{aaaa}=H_{1}(t; \bm{\theta}_{0})+o(\frac{1}{\sqrt{n}})-H_{2}(t; \bm{\theta}_{0})+o(\frac{1}{\sqrt{n}})+o(\frac{1}{\sqrt{n}}) \\
&\phantom{aaaa}=H_{1}(t; \bm{\theta}_{0})-H_{2}(t; \bm{\theta}_{0})+o(\frac{1}{\sqrt{n}}).
\end{split}
\end{equation}

Hence, $\sqrt{n}(\tilde{G}-\hat{G})$ can be approximated by $\sqrt{n}(H_{1}-H_{2}-\hat{G})$ uniformly in $t$. It is left to show that the finite dimension distribution of $\sqrt{n}(H_{1}-H_{2}-\hat{G})$ converges and $\sqrt{n}(H_{1}-H_{2}-\hat{G})$ is tight.
By the same derivation in the proof of the Lemma 3.5 in \cite{Lu2007}, $\forall k$ and $t_{1}, \dots, t_{k}$,
\begin{equation}
\sqrt{n}(H_{1}(t_1)-H_{2}(t_1)-\hat{G}(t_1), \dots, H_{1}(t_k)-H_{2}(t_k)-\hat{G}(t_k))^{T}\stackrel{d}{\to}N(\bm{0}, \bm{\Delta})
\end{equation}
where $\bm{\Delta}$ is the covariance matrix depending on the covariance function
\begin{equation}
\begin{split}
&\mathrm{Cov}(\sqrt{n}(H_{1}(s)-H_{2}(s)-\hat{G}(s)), \sqrt{n}(H_{1}(t)-H_{2}(t)-\hat{G}(t))) \\
=&\sum_{j=0}^{m}\rho_{j}\sum_{j=1}^{m}\rho_{j}A_{j}(s\wedge t)-(\bar{\bm{A}}^{T}(t)\bm{\rho}, \bar{\bm{B}}^{T}(t)\bar{\bm{\rho}})\bm{S}^{-1}(\bar{\bm{A}}^{T}(s)\bm{\rho}, \bar{\bm{B}}^{T}(s)\bar{\bm{\rho}})^{T}.
\end{split}
\end{equation}

Also, the tightness of $\sqrt{n}(H_{1}-\hat{G})$ and $\sqrt{n}H_{2}$ can be proved by Lemma 3.6 and Lemma 3.7 in \cite{Lu2007}. Therefore, we have shown that $\sqrt{n}(\tilde{G}-G)$ converges weakly to a zero-mean Gaussian process in $D[-\infty, \infty]$. By the same derivation in the proof of the Theorem 3.9 in \cite{Lu2007}, we finally have
\begin{equation}
\sqrt{n}(\tilde{G}(t)-G(t))\stackrel{d}{\to}N(0, \sigma(t)),
\end{equation}
where 
\begin{equation}
\begin{split}
\sigma(t)=&\bigg(\sum_{j=0}^{m}\rho_{j}\bigg)\bigg(G(t)-G^{2}(t)-\sum_{j=1}^{m}\rho_{j}A_{j}(t)\bigg) \\
&+(\bar{\bm{A}}^{T}(t)\bm{\rho}, \bar{\bm{B}}^{T}(t)\bar{\bm{\rho}})\bm{S}^{-1}(\bar{\bm{A}}^{T}(t)\bm{\rho}, \bar{\bm{B}}^{T}(t)\bar{\bm{\rho}})^{T}.
\end{split}
\label{Variance of hat(G)}
\end{equation}
\end{appendices}
\clearpage
\bibliographystyle{siam}
\bibliography{Radon}

\begin{thebibliography}{10}

\bibitem{PA.GOV}
\url{https://www.dep.pa.gov/Business/RadiationProtection/RadonDivision/Pages/Radon-in-the
  home.aspx}.
\newblock {PA Department of Environmental Protection}.

\bibitem{EPA2003}
{\em {EPA Assessment of Risks from Radon in Homes: EPA 402-R-03-003}}.
\newblock
  \url{https://www.epa.gov/sites/production/files/2015-05/documents/402-r-03-003.pdf}.
\newblock {U.S. Environmental Protection Agency}.

\bibitem{WHO2009}
{\em Indoor radon: A public health perspective}, Int. J. Environ. Stud, 67
  (2009), p.~108.
\newblock World Health Organization.

\bibitem{SDF}
{\sc K.~Benjamin, O.~V. De, and S.~Michael}, {\em Statistical Data Fusion},
  World Scientific, 2017.

\bibitem{CaseyEtAl2015}
{\sc J.~A. Casey, E.~L. Ogburn, S.~G. Rasmussen, J.~K. Irving, J.~Pollak, P.~A.
  Locke, and B.~S. Schwartz}, {\em {Predictors of indoor radon concentrations
  in Pennsylvania, 1989--2013}}, Environmental health perspectives, 123 (2015),
  pp.~1130--1137.

\bibitem{Fokianos2004}
{\sc K.~Fokianos}, {\em Merging information for semiparametric density
  estimation}, Journal of the Royal Statistical Society: Series B (Statistical
  Methodology), 66 (2004), pp.~941--958.

\bibitem{FKQS}
{\sc K.~Fokianos, B.~Kedem, J.~Qin, and D.~A. Short}, {\em A semiparametric
  approach to the one-way layout}, Technometrics, 43 (2001), pp.~56--65.

\bibitem{DHHS2012}
{\sc S.~Keith, J.~Doyle, C.~Harper, M.~Mumtaz, O.~Tarrago, D.~Wohlers,
  G.~Diamond, M.~Citra, and L.~Barber}, {\em Toxicological profile for radon},
  (2012).

\bibitem{Lu2007}
{\sc G.~Lu}, {\em {Asymptotic Theory for Multiple-Sample Semiparametric Density
  Ratio Models and its Application to Mortality Forecasting}}, (2007).
\newblock Ph.D. Dissertation, University of Maryland, College Park.

\bibitem{Qin2017}
{\sc J.~Qin}, {\em Biased Sampling, Over-identified Parameter Problems and
  Beyond}, Springer, 2017.

\bibitem{QinLawless}
{\sc J.~Qin and J.~Lawless}, {\em Empirical likelihood and general estimating
  equations}, The Annals of Statistics,  (1994), pp.~300--325.

\bibitem{QinZhang97}
{\sc J.~Qin and B.~Zhang}, {\em A goodness-of-fit test for logistic regression
  models based on case-control data}, Biometrika, 84 (1997), pp.~609--618.

\bibitem{Voulgaraki2011}
{\sc A.~Voulgaraki}, {\em {Semiparametric Regression and Mortality Rate
  Prediction}}, (2011).
\newblock Ph.D. Dissertation, University of Maryland, College Park.

\bibitem{VoulgarakiEtAl2012}
{\sc A.~Voulgaraki, B.~Kedem, B.~I. Graubard, et~al.}, {\em Semiparametric
  regression in testicular germ cell data}, The Annals of Applied Statistics, 6
  (2012), pp.~1185--1208.

\bibitem{Zhang2000b}
{\sc B.~Zhang}, {\em A goodness of fit test for multiplicative-intercept risk
  models based on case-control data}, Statistica Sinica,  (2000), pp.~839--865.

\end{thebibliography}
\end{document}